\documentclass{emulateapj}
\usepackage{apjfonts}

\lefthead{HAN} \righthead{SOURCE COMPANIONS AND OPTICAL DEPTH}

\begin{document}
\title{Effect of Binary Source Companions on the Microlensing Optical 
Depth Determination toward the Galactic Bulge Field}

\author{Cheongho Han}
\affil{Department of Physics, Institute for Basic Science
Research, Chungbuk National University, Chongju 361-763, Korea;
cheongho@astroph.chungbuk.ac.kr}


\begin{abstract}

Currently, gravitational microlensing survey experiments toward the 
Galactic bulge field utilize two different methods of minimizing 
blending effect for the accurate determination of the optical depth 
$\tau$.  One is measuring $\tau$ based on clump giant (CG) source 
stars and the other is using `Difference Image Analysis (DIA)' 
photometry to measure the unblended source flux variation.  Despite 
the expectation that the two estimates should be the same assuming 
that blending is properly considered, the estimates based on CG stars
systematically fall below the DIA results based on all events with 
source stars down to the detection limit.  Prompted by the gap, we 
investigate the previously unconsidered effect of companion-associated 
events on $\tau$ determination.  Although the image of a companion 
is blended with that of its primary star and thus not resolved, the 
event associated with the companion can be detected if the companion 
flux is highly magnified.  Therefore, companions work effectively as 
source stars to microlensing and thus neglect of them in the source 
star count could result in wrong $\tau$ estimation.  By carrying out 
simulations based on the assumption that companions follow the same 
luminosity function of primary stars, we estimate that the contribution 
of the companion-associated events to the total event rate is
$\sim 5f_{\rm bi}\%$ for current surveys and can reach up to 
$\sim 6f_{\rm bi}\%$ for future surveys monitoring fainter stars, 
where $f_{\rm bi}$ is the binary frequency.  Therefore, we conclude 
that the companion-associated events comprise a non-negligible fraction 
of all events. However, their contribution to the optical depth is not 
large enough to explain the systematic difference between the optical 
depth estimates based on the two different methods.

\end{abstract}

\keywords{binaries: general -- gravitational lensing}

\section{Introduction}

The value of the microlensing optical depth toward the Galactic bulge 
field has been evolved with the refinement of both experimental 
measurements and theoretical predictions.  \citet{paczynski91} and 
\citet{griest91} first predicted the optical depth of $\tau\sim 0.4
-0.5 \times 10^{-6}$ assuming all events were caused by known disk 
stars.  However, the first measurement of the optical depth of 
$\tau=3.3 \times 10^{-6}$ reported by OGLE \citep{udalski94} was 
significantly higher than the earlier predictions.  This prompted 
\citet{kiraga94a} to evaluate the lens contribution of bulge stars 
in addition to disk stars.  Soon after, optical depths based on 
more refined models of the Galactic bulge were suggested by a number 
of authors \citep{kiraga94b, zhao95, han95, metcalf95, zhao96, 
bissantz97, peale98, evans02}.  The values of the new predictions  
were in the range of $\tau\sim 0.8-2.0\times 10^{-6}$, but these 
values still fall systematically below the measurements by OGLE and 
MACHO \citep{alcock97}.

To reconcile the predictions and measurements, various explanations were 
suggested.  A set of these explanations suggested new populations of 
events.  \citet{mollerach96} pointed out non-negligible contribution 
of disk self-lensing events.  \citet{nair99} indicated that disk source 
stars behind the bulge has much higher optical depth than bulge stars, 
and thus although they are a very small fraction of the stars in the 
Baade's window, they can contribute $\sim 5\%-10\%$ of the optical 
depth.  \citet{cseresnjes01} pointed out similar contribution of events 
associated with source stars belonging to the Sagittarius dwarf galaxy.  
\citet{sevenster01} mentioned the possibility of events caused by lenses 
in an inner ring located roughly halfway between the observer and bulge.

Another set of explanations pointed out potential biases in the optical 
depth measurements.  One of such biases is the the ``magnification-bias 
effect'' \citep{nemiroff94}, pointed out by \citet{alard97} and 
\citet{han97}.  The bulge is very crowded field and thus the images of 
stars suffer from severe blending.  For efficient processing of images 
taken toward such a crowded field, earlier lensing experiments registered 
bright resolved stars on a template image and monitored the brightness 
variation of only these stars.  For such experiments, although a source 
star is fainter than the detection limit imposed by crowding and thus 
not registered, it is still possible to detect an event if the star is 
located close to the seeing disk of a bright registered star. The 
optical depth is measured based on the number of monitored stars, 
$N_\star$, by
\begin{equation}
\tau={\pi\over 2N_\star T} \sum_{i=1}^{N_{\rm tot}}{t_{{\rm E},i} 
\over \epsilon(t_{{\rm E},i})},
\label{eq1}
\end{equation}
where $T$ is the total observation time, $N_{\rm tot}$ is the total 
number of detected events, $t_{{\rm E},i}$ are the Einstein timescales 
of the individual events, and $\epsilon(t_{{\rm E},i})$ is the detection 
efficiency of events as a function of $t_{\rm E}$.  Then, if these 
faint source stars are not taken into consideration, the number of 
source stars effectively monitored by the survey is systematically 
underestimated, and thus the optical depth is overestimated.

Observational efforts to refine the optical depth measurements were 
focused also on blending.  One simple but efficient method for the 
accurate determination of the optical depth was measuring $\tau$
based on events associated with only clump giant (CG) source stars, 
because these bright bulge stars are not strongly affected by blending.  
Four optical depth measurements based on CG-associated events were 
reported by \citet{popowski01}, \citet{afonso03}, \citet{popowski04}, 
and \citet{sumi05}.  These values range $\tau\sim 0.9-2.2\times 10^{-6}$, 
approaching close to the theoretical predictions.  Another effort of 
minimizing the blending effect was the adoption of a new photometry 
technique that measures the source star flux variation on images 
obtained by subtracting observed images from a seeing-convolved 
reference image \citep{tomaney96, alard98, alcock99, wozniak00, alard00, 
bond01}.  By suing the DIA method, one can improve the photometry of 
events because the measured flux variation is not affected by blending.
The optical depth measurements by using this so called ``Difference 
Image Analysis'' (DIA) method based on all events with source stars 
down to the detection limit were published by \citet{alcock00} and 
\citet{sumi03}.  These estimates range $\tau\sim 3.2-3.4 \times 10^{-6}$, 
and thus they are systematically greater than the estimates based on CG 
stars and the theoretical predictions.

Prompted by the difference between the optical depth estimates based 
on CG stars and the DIA result, we investigate the effect of unresolved 
binary companions to source stars on the microlensing optical depth 
determination.  Due to the characteristics of the DIA photometry, a 
significant fraction of events are associated with faint source stars
which could not be identified by the traditional DoPHOT \citep{schechter93}.  
Therefore, estimation of the optical depth based on the DIA photometry 
requires consideration of faint stars effectively working as source stars 
to lensing.  In the previous DIA estimates, this has been done based on 
a luminosity function (LF) obtained by using high-resolution instruments 
such as the {\it Hubble Space Telescope} ({\it HST}).  However, companions 
to source stars cannot be resolved even with the {\it HST}, and thus the 
contribution of the companions to the effective number of source stars 
has {\it not} been taken into consideration in the previous optical 
depth estimations.  Considering that majority of stars have companions 
\citep{abt76, abt83, duquennoy91}, the fraction of events associated 
with companion stars might not be negligible and could explain a 
significant fraction of the gap between the optical depth estimates 
based on CG stars and the DIA results.

\section{Companion-Associated Events}

Let us think about an event occurred on a companion star whose image is 
blended with that of its brighter primary star.  The separation between 
the primary and companion stars is, in general, much larger than the 
Einstein ring radius of a typical Galactic bulge event caused by a 
low-mass star of
\begin{equation}
r_{\rm E}\sim 1.9\ {\rm AU} \left( {M\over 0.3\ M_\odot}\right)^{1/2}
\left( {D_L\over 6\ {\rm kpc}}\right)^{1/2} 
\left( 1-{D_L\over D_S}\right)^{1/2},
\label{eq2}
\end{equation}
where $M$ is the mass of the lens, and $D_L$ and $D_S$ are the distances 
to the lens and source, respectively.  Then, the primary star, in most 
cases, is not participating in the lensing magnification and thus it 
simply works as a blending source \citep{dominik98, han98}.  In this 
case, the apparent magnification of the event is 
\begin{equation}
A_{\rm obs}(u) = {F_{\rm p}+A(u)F_{\rm c}\over F_{\rm p}+F_{\rm c}} 
= {F_{\rm p}/F_{\rm c}+A(u)\over F_{\rm p}/F_{\rm c} +1},
\label{eq3}
\end{equation}
where $A(u)$ is the true magnification of the event occurred on the 
companion star, $u$ is the lens-source (companion star) angular 
separation in units of the angular Einstein ring radius $\theta_{\rm E}
=r_{\rm E}/D_L$, and $F_{\rm p}$ and $F_{\rm c}$ are the fluxes of the 
primary and companion stars, respectively.  The true magnification is 
related to the normalized lens-source separation $u$ by
\begin{equation} A= {u^2+2 \over u(u^2+4)^{1/2}}.
\label{eq4}
\end{equation}

Although the image of a companion star is blended, the event associated 
with the companion star can be detected if its flux is magnified highly 
enough to make the apparent magnification of the combined flux of the
 primary and companion higher than a threshold value. Due to blending, 
the threshold magnification for the event detection becomes higher.  
The increase threshold magnification is  
\begin{equation}
{A'}_{\rm th}=A_{\rm th} \left( 1+{F_{\rm p}\over F_{\rm c}} \right)
-{F_{\rm p}\over F_{\rm c}},
\label{eq5}
\end{equation}
where $A_{\rm th}$ is the threshold magnification without blending.  Then, 
the threshold lens-source separation (impact parameter) corresponding to 
the increased threshold magnification is 
\begin{equation}
{u'}_{0,{\rm th}} = \left[ {2\over \left( 1-{A'}_{\rm th}^{-2} 
\right)^{1/2}}  -2 \right]^{1/2},
\label{eq6}
\end{equation}
implying that as long as the lens approaches the companion star closer 
than ${u'}_{0,{\rm th}}$, the event associated with the companion can 
be detected.  Therefore, companions can work effectively as source stars 
to lensing and thus the events associated with them should be taken into 
consideration in the optical depth determination.

\section{Companion Contribution to Source Stars}

Then, how much is the contribution of events associated with companion 
stars to the total optical depth, $\tau_{\rm c}$, toward the Galactic 
bulge field.  For this estimation, we compute the ratio of the number 
of events associated with companions, $N_{\rm c}$, to the total number 
of events, $N_{\rm tot}$.  The optical depth is related to the mass 
distribution along the line of sight toward the observation field, while 
the event rate is additionally dependent on the mass function of the lens 
matter.  However, since the lens does not discriminate whether the source 
star is primary or companion, the optical depth contribution of the 
companion-associated events is equivalent to their contribution to the 
total event rate, i.e.\ $\tau_{\rm c}/\tau=N_{\rm c}/N_{\rm tot}$.

\begin{figure}[tb]
\epsscale{1.15}
\plotone{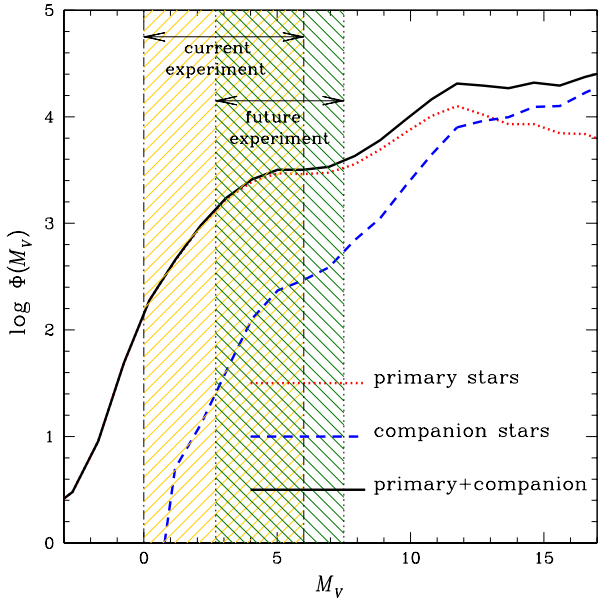}
\caption{\label{fig:one}
Luminosity of function of source stars effectively participating in 
microlensing (effective luminosity function) toward the Galactic bulge 
field.  The dotted and and dashed curves represent the LFs of the 
primary and companion stars, respectively, and the solid curve is the 
combined LF.  The effective LF of the companion source stars is for 
the case of $f_{\rm bi}=1.0$, where $f_{\rm bi}$ is the binary frequency.  
The LFs are arbitrarily normalized.  The shaded regions represent the 
source stars brightness ranges of the current and future lensing surveys.
}\end{figure}

\begin{deluxetable}{lcc}
\tabletypesize{\scriptsize}
\tablecaption{Contribution of Companion-Associated Events\label{table1}}
\tablewidth{0pt}
\tablehead{
\colhead{surveys} &
\colhead{source brightness range}  &
\colhead{$\tau_{\rm c}/\tau$} 
}
\startdata
\smallskip
current-type        & $0.0\lesssim M_V\lesssim 6.0$  & $\sim 0.047f_{\rm bi}$ \\
future space-based  & $2.7\lesssim M_V\lesssim 7.5$  & $\sim 0.060f_{\rm bi}$ \\
\enddata
\end{deluxetable}

We estimate $\tau_{\rm c}/\tau$ as follows.  First, we assume that 
companions follow the same LF as that of primary stars \citep{duquennoy91}.  
With this assumption, we assign the brightnesses of the primary and 
companion source stars based on the binary-corrected $V$-band LF of 
stars in the solar neighborhood listed in \citet{allen00}.\footnote{For 
the source star brightness, it might be thought that using the LF of 
Galactic bulge stars from {\it HST} observations, e.g.\ \citet{holtzman98},  
would be a better choice.  However, this LF is not binary corrected 
because companions cannot be resolved even with the {\it HST}.  Since 
the purpose of this paper is investigating the effect of the source 
companion to the optical depth determination, we use the binary-corrected 
LF, although it is based on stars in the solar neighborhood. We note, 
however, that there is no significant evidence for the variation of 
the field of LF from place to place.} Once each pair of the primary 
and companion stars is produced, we then calculate the threshold 
magnification ${A'}_{\rm th}$ required for the event detection  and 
the corresponding threshold impact parameter ${u'}_{0,{\rm th}}$.  
The number of events having magnifications higher than $A'_{th}$ is 
directly proportional to $u'_{0,{\rm th}}$ because the distribution of 
events is uniform in impact parameter.  Then, the contribution of the 
companion-associated events to the total event rate is computed by 
\begin{equation}
{N_{\rm c}\over N_{\rm tot}} = {N_{\rm c}\over N_{\rm p}+N_{\rm c}},
\label{eq7}
\end{equation}
where 
\begin{equation}
N_{\rm p}=\sum_{i=1}^{N_{\star,{\rm p}}}u_{0,{\rm th,p}},
\label{eq8}
\end{equation}
\begin{equation}
N_{\rm c}=f_{\rm bi}\sum_{i=1}^{N_{\star,{\rm p}}}u_{0,{\rm th,c}},
\label{eq9}
\end{equation}
$N_{\star,{\rm p}}$ is the total number of monitored source primaries that 
would have produced events in the absence of binarity, $f_{\rm bi}$ is the 
binary frequency, and  $u_{0,{\rm th,p}}$ and $u_{0,{\rm th,c}}$ represent 
the threshold impact parameters of the events associated with the primary 
and companion, respectively. We note that the event associated with the 
primary star is also affected by blending caused by the flux from 
the companion star, and thus $u_{0,{\rm th,p}}\leq 1.0$ and 
$N_{\rm p}\leq N_{\star,{\rm p}}$.  For the computation of the 
threshold impact parameter, we assume that events are detected when 
$A_{\rm obs}\geq 3/\sqrt{5}$ adopting the conventional threshold 
magnification. We also assume that the combined (primary plus companion) 
brightness of the source star monitored by the current surveys is in 
the range of $0.0\lesssim M_V \lesssim 6.0$.  With the distance 
modulus of $V-M_V\sim 14.5$ and extinction of $A_V\sim 1.0$, this 
range corresponds to the apparent magnitude range of 
$15.5\lesssim V\lesssim 21.5$.  As the brightness of the monitoring 
source star decreases, the relative companion/primary flux ratio 
increases, and thus the rate of the companion-associated events 
increases.  The future space-based lensing survey using the 
{\it Galactic Exoplanet Survey Telescope} ({\it GEST}) mission 
proposed by \citet{bennett02} plans to monitor faint main-sequence 
stars to optimize the detections of terrestrial planets by minimizing 
finite-source effect.  To examine the dependence on the source star 
brightness, we also estimate the fraction of the companion-associated 
events from this future survey to examine the effect of source star 
brightness.  For this estimation, we assume that the range of the 
source star brightness is $2.7\lesssim M_V \lesssim 7.5$, which 
corresponds to early F to late K-type main-sequence stars.

In Tabel~\ref{table1}, we present the estimated contribution of the 
companion-associated events to the total event rate expected from 
the current and future lensing surveys.  In Figure~\ref{fig:one}, 
we also present the LFs of the primary and companion source stars 
effectively participating in lensing ({\it effective luminosity 
function}), where the companion LF is for the case $f_{\rm bi}=1.0$ 
to show the possible maximum contribution of the companion-associated 
events.  The result says that the contribution of the 
companion-associated events is $N_{\rm c}/N_{\rm tot}\sim 5f_{\rm bi}\%$ 
for the current lensing surveys and it is slightly higher values of 
$\sim 6f_{\rm bi}\%$ for the future lensing survey.  Therefore, 
these events do contribute to the total bulge event rate and thus 
to the optical depth.  However, the contribution is not large enough 
to explain an important fraction of the gap between the optical depth 
estimates based on CG stars and the DIA results, implying that exploring 
other possible reasons that can explain the gap is needed.


\section{Conclusion}

Prompted by the gap between the microlensing optical depth estimates 
based on CG stars and the DIA result, we investigated the previously 
unconsidered effect of unresolved binary companions to source stars 
on the determination of the optical depth toward the Galactic bulge 
field.  By carrying out simulations based on the assumption that 
companions follow the same LF of primary stars, we assessed the 
contribution of the companion-associated events to the total optical 
depth would be $\tau_{\rm c}/\tau\sim 5f_{\rm bi}\%$ for current 
lensing surveys and can reach up to $\sim 6f_{\rm bi}\%$ for future 
surveys monitoring fainter stars.  We, therefore, conclude that events 
associated with companion stars comprises a non-negligible fraction of 
the total events, but their contribution to the optical depth is not 
large enough to explain the systematic difference between the optical 
depth estimates based on the two different methods.

\acknowledgments 
This work was supported by the Astrophysical Research Center for the 
Structure and Evolution of the Cosmos (ARCSEC") of Korea Science \& 
Engineering Foundation (KOSEF) through Science Research Program (SRC) 
program.

\end{document}